\documentclass[aps,prd,preprint,superscriptaddress]{revtex4}
\usepackage{amsmath,amssymb,multirow}
\usepackage{graphicx}
\usepackage[hidelinks]{hyperref}
\usepackage{placeins}
\usepackage{float}

\begin{document}

\title{M1 Radiative Transitions in the $B_c$ System after the $B_c^\ast(1S)$ Observation: Radial-Node Cancellation and Quark-Line Filtering}

\author{Bing-Dong Wan}
\email{wanbd@lnnu.edu.cn}
\affiliation{School of Physics and Electronic Technology, Liaoning Normal University, Dalian 116029, China}
\affiliation{Center for Theoretical and Experimental High Energy Physics, Liaoning Normal University, Dalian 116029, China}

\begin{abstract}
The recent ATLAS observation of the $B_c^*(1S)$ meson fixes the photon momentum for $B_c^*(1S)\to B_c(1S)\gamma$. We calculate four $1S$ and $2S$ M1 radiative transitions with relativistic Salpeter wave functions calibrated to the measured ground-state masses, and obtain $\Gamma[B_c^*(1S)\to B_c(1S)\gamma]=5.24\times10^{-2}$~keV. For the hindered $2S\to1S$ transitions, contributions from different radial-momentum regions largely cancel, leaving form factors about 25 times smaller than that of the ground-state allowed transition. The cancellation factors, defined as the ratio of the integral of the absolute radial contribution to the magnitude of the net integral, are approximately 18--20. Nevertheless, the hindered radiative widths are not suppressed, because their much larger photon momenta compensate for the smaller form factors. Separating the photon-emission contributions of the $b$ quark and $\bar c$ antiquark, we find particularly strong cancellation in the $b$ contribution to $B_c(2S)\to B_c^*(1S)\gamma$: only about $0.46\%$ of the integrated absolute radial contribution remains after cancellation, compared with about $4.5\%$ for the $\bar c$ contribution. The unmeasured $B_c^*(2S)$ mass remains the main limitation on the $2S\to2S$ radiative width.
\end{abstract}

\maketitle

\section{Introduction}
\label{sec:intro}

The $B_c$ meson, composed of a bottom quark and a charm antiquark ($b\bar c$), occupies a unique position in heavy-quark spectroscopy~\cite{Eichten:2019gig,Ding:2021dwh}.
Unlike the equal-mass charmonium ($c\bar c$) and bottomonium ($b\bar b$) systems, the $B_c$ system carries open flavor, so the ground-state $B_c$ meson cannot annihilate either strongly or electromagnetically and therefore decays weakly. Excited $B_c$ states below the open-flavor threshold can, however, undergo electromagnetic or hadronic transitions to lower $B_c$ states.
Because the two constituents carry different flavors, the $B_c$ meson is not an eigenstate of charge conjugation, and $C$ parity is therefore not a good quantum number, while the different electric charges of the $b$ and $\bar c$ quarks lead to nontrivial interference between their electromagnetic currents in radiative transitions.
These features make the $B_c$ system an ideal laboratory for testing relativistic bound-state dynamics in QCD~\cite{Eichten:1994gt,Godfrey:2004ya}.
We work with the $B_c^-=(b\bar c)$ convention throughout; the charge-conjugate $B_c^+=(c\bar b)$ transition has the same radiative width.

CMS~\cite{CMS:2019uhm} and LHCb~\cite{LHCb:2019bem} observed two structures in the $B_c(2S)$ region, with a peak separation of $\Delta_{\rm peak}=29.1\pm1.5_{\rm stat}\pm0.7_{\rm syst}$~MeV and approximately $31$~MeV, respectively.
More recently, the ATLAS collaboration~\cite{ATLAS:2026ubk} reported the first direct observation of the $B_c^{*+}(1S)$ meson with a significance exceeding $8\sigma$, measuring the ground-state hyperfine splitting
\begin{equation}
\Delta_{\rm HF}(1S)=M[B_c^*(1S)]-M[B_c(1S)]=64.5\pm1.4_{\rm stat}{}^{+1.0}_{-1.4\,\rm syst}~\text{MeV}.
\label{eq:delta_hf}
\end{equation}
This measurement fixes the $B_c^*$ mass at approximately $6339$~MeV and therefore determines the phase space for the $B_c^*\to B_c\gamma$ transition.

The magnetic dipole (M1) radiative transition $B_c^*\to B_c\gamma$ is the dominant decay mode of the $B_c^*$ meson, and its partial width $\Gamma\propto k_\gamma^3|F(0)|^2$ carries direct information about the $b\bar c$ bound-state wave function through the transition form factor $F(0)$.
Prior to the ATLAS measurement, the unknown $B_c^*$ mass introduced an additional uncertainty in the photon momentum $k_\gamma$, on top of the model dependence of the transition form factor. Lattice QCD studies~\cite{Gregory:2009hq,Mathur:2018epb} predicted $M[B_c^*(1S)]$ in the range $6320$--$6340$~MeV, consistent with the ATLAS result, but the experimental determination now eliminates the need for model-dependent mass inputs.
With the $B_c^*(1S)$ mass now experimentally determined, the dominant model-dependent kinematic ambiguity is removed, and the remaining model-dependent uncertainty in the radiative width is primarily associated with $|F(0)|^2$---a quantity that directly probes the relativistic structure of the $b\bar c$ wave function.

The M1 transitions of the $B_c$ system have been studied in various theoretical frameworks.
Ke {\it et al.}~\cite{Ke:2010vx} performed the first systematic Bethe-Salpeter (BS) calculation of $B_c$ M1 transitions, giving predictions for $B_c^*(1S)\to B_c(1S)\gamma$, $B_c^*(2S)\to B_c(1S)\gamma$, and $B_c(2S)\to B_c^*(1S)\gamma$.
Wang {\it et al.}~\cite{Wang:2022cxy} subsequently provided the complete Salpeter wave functions for the $B_c$ system, including the full relativistic partial-wave content of the $0^-$ and $1^-$ states.
Complementary strategies for identifying the $B_c^*$ experimentally have also been proposed using angular-momentum and kinematic conservation laws~\cite{Liu:2022bdq}.
The radiative transitions of the $B_c$ system have also been investigated in nonrelativistic potential models~\cite{Li:2019qkt,Asghar:2019abc}, a constituent-quark model~\cite{MartinGonzalez:2022bc}, the Cornell potential model~\cite{Gao:2024yvz}, the light-front quark model~\cite{Ridwan:2024ngc}, and a screened-potential framework~\cite{Patel:2026prg}.
The general understanding of allowed and hindered M1 transitions in heavy quarkonium has been developed through effective field theory~\cite{Brambilla:2005zw} and covariant light-front approaches~\cite{Wang:2010npa}.
Related node-induced suppression and filtering effects in open-charm decays of excited charmonia have been studied within the Bethe--Salpeter framework~\cite{Wan:2026eek,Wan:2026csm}.

Building on earlier Bethe--Salpeter studies of $B_c$ M1 transitions~\cite{Ke:2010vx}, the present analysis differs in two respects. First, the measured $B_c^*(1S)$ mass fixes the kinematics of the ground-state radiative transition. Second, besides computing the complete form factors and widths, we examine how different radial-momentum regions contribute to the transition form factors. For the hindered channels, contributions of opposite signs largely cancel. We also separate the contributions from photon emission by the $b$ quark and the $\bar c$ antiquark to identify which term is more strongly affected by this cancellation. This analysis clarifies the role of the radial nodes and the unequal-mass recoil kinematics in the M1 amplitudes.

The paper is organized as follows.
Section~\ref{sec:formalism} introduces the instantaneous Bethe--Salpeter framework, the interaction kernel, the Salpeter wave functions, and the M1 transition current.
Section~\ref{sec:wavefunctions} presents the mass calibration and the resulting Salpeter wave functions, including the $1S/2S$ radial structure.
Section~\ref{sec:results} gives the computed M1 form factors and widths for all four channels under both model-consistent and experiment-informed kinematics.
Section~\ref{sec:phasespace} quantifies the separation of kinematics and bound-state dynamics.
Section~\ref{sec:hierarchy} analyzes the radial-node cancellation and the quark-line-resolved filtering of the hindered transitions.
Section~\ref{sec:summary} summarizes our findings.

\section{Bethe--Salpeter Framework and M1 Transition Current}
\label{sec:formalism}

\subsection{Instantaneous Bethe--Salpeter equation and interaction kernel}

The relativistic $q\bar q$ bound state is described by the Bethe--Salpeter (BS) equation~\cite{Salpeter:1951sz}. For a meson with total momentum $P$, relative momentum $q$, quark momentum $p_1=\alpha_1P+q$, and antiquark momentum $p_2=\alpha_2P-q$, with $\alpha_i=m_i/(m_1+m_2)$, we write
\begin{equation}
\chi_P(q)=iS_1(p_1)\int\frac{d^4k}{(2\pi)^4}\,V(P;k,q)\,\chi_P(k)\,S_2(-p_2),
\label{eq:bse}
\end{equation}
where $\chi_P(q)$ is the four-dimensional BS amplitude and $V$ is the interaction kernel. Following the instantaneous formulation of Refs.~\cite{Salpeter:1952ib,Chang:2003ua,Chang:2004im}, we decompose
\begin{equation}
q^\mu=q_P\frac{P^\mu}{M}+q_\perp^\mu,\qquad q_P=\frac{P\cdot q}{M},\qquad P\cdot q_\perp=0,
\end{equation}
and take the kernel to depend only on the transverse three-momenta, $V(P;k,q)\to V(k_\perp,q_\perp)$. The corresponding three-dimensional Salpeter wave function is defined as
\begin{equation}
\varphi_P(q_\perp)=i\int\frac{dq_P}{2\pi}\,\chi_P(q).
\end{equation}
After the relative-energy integration, the quark and antiquark propagators are separated into positive- and negative-energy parts. With $\omega_i=\sqrt{m_i^2+\mathbf q^2}$, the corresponding projectors can be written as
\begin{equation}
\Lambda_i^\pm(q_\perp)=\frac{1}{2\omega_i}
\left[\frac{\not{P}}{M}\omega_i\pm\left(\not{q}_\perp+J_i m_i\right)\right],
\qquad J_1=+1,\quad J_2=-1.
\end{equation}
With these definitions, the BS equation can be written as
\begin{equation}
\begin{aligned}
\chi_P(q)&=S_1(p_1)\,\eta_P(q_\perp)\,S_2(-p_2),\\
\eta_P(q_\perp)&=\int\frac{d^3k_\perp}{(2\pi)^3}\,
V(q_\perp,k_\perp)\varphi_P(k_\perp),
\end{aligned}
\label{eq:eta_def}
\end{equation}
where $\eta_P$ is the three-dimensional interaction term. Introducing the projected Salpeter amplitudes,
\begin{equation}
\varphi_P^{\pm\pm}=\Lambda_1^\pm\frac{\not{P}}{M}
\varphi_P\frac{\not{P}}{M}\Lambda_2^\pm,
\end{equation}
one obtains the coupled complete Salpeter equations,
\begin{align}
(M-\omega_1-\omega_2)\varphi_P^{++}&=\Lambda_1^+\eta_P\Lambda_2^+,\\
(M+\omega_1+\omega_2)\varphi_P^{--}&=-\Lambda_1^-\eta_P\Lambda_2^- ,
\end{align}
together with the Salpeter constraints $\varphi_P^{+-}=\varphi_P^{-+}=0$.
We solve these complete Salpeter equations to obtain the bound-state masses and radial functions, while the electromagnetic transition current below retains the positive-energy projected amplitudes $\varphi^{++}$, following Ref.~\cite{Ke:2010vx}.

We adopt the screened Cornell-type interaction kernel used in the Salpeter formulation of Ref.~\cite{Kim:2003ny}. In momentum space, with $\boldsymbol{\Delta}=\mathbf k-\mathbf q$, it is written as
\begin{equation}
V(\boldsymbol{\Delta})=(2\pi)^3V_s(\boldsymbol{\Delta})
+\gamma^0\otimes\gamma^0\,(2\pi)^3V_v(\boldsymbol{\Delta}),
\label{eq:kernel}
\end{equation}
with
\begin{align}
V_s(\boldsymbol{\Delta})&=-\left(\frac{\lambda}{\alpha}+V_0\right)\delta^3(\boldsymbol{\Delta})
+\frac{\lambda}{\pi^2}\frac{1}{(\boldsymbol{\Delta}^2+\alpha^2)^2},\\
V_v(\boldsymbol{\Delta})&=-\frac{2}{3\pi^2}\frac{\alpha_s(\boldsymbol{\Delta})}{\boldsymbol{\Delta}^2+\alpha^2},\\
\alpha_s(\boldsymbol{\Delta})&=\frac{12\pi}{27}\,
\frac{1}{\ln\!\left(a+\boldsymbol{\Delta}^2/\Lambda_{\rm QCD}^2\right)}.
\label{eq:running_alpha}
\end{align}
Here $V_s$ represents the screened scalar confinement interaction and
$V_v$ the time-like vector one-gluon-exchange interaction. The running
coupling in the vector interaction follows the effective parametrization
used in the present Salpeter kernel. The infrared regulator $\alpha$
implements the screening prescription. The numerical values of the
physical kernel parameters are summarized in Appendix~A; following the
phenomenological Salpeter prescription, the additive
constant $V_0$ is calibrated separately in the $0^-$ and $1^-$ sectors to
reproduce the measured ground-state masses, as described in Sec.~\ref{sec:wavefunctions}.

\subsection{Salpeter wave functions for the \texorpdfstring{$0^-$}{0-} and \texorpdfstring{$1^-$}{1-} states}

The covariant wave-function structures follow Ref.~\cite{Wang:2022cxy}. For a pseudoscalar state, we use
\begin{equation}
\varphi_{0^-}(q_\perp)=\left[g_1\not{P}+g_2M+g_3\not{q}_\perp
+g_4\frac{\not{P}\not{q}_\perp}{M}\right]\gamma_5,
\label{eq:phi0_full}
\end{equation}
where the radial functions depend on $\mathbf q^2=-q_\perp^2$. The Salpeter constraints leave two independent functions, chosen as $g_1$ and $g_2$; the remaining two are fixed by
\begin{equation}
g_3=\frac{g_2M(-\omega_1+\omega_2)}{m_2\omega_1+m_1\omega_2},\qquad
g_4=-\frac{g_1M(\omega_1+\omega_2)}{m_2\omega_1+m_1\omega_2}.
\end{equation}
The structures proportional to $g_1$ and $g_2$ constitute the dominant ${}^1S_0$ component, while the momentum-dependent terms provide the relativistic $P$-wave admixture.

For a vector state with polarization $\epsilon^\mu$, the general instantaneous amplitude is
\begin{align}
\varphi_{1^-}(q_\perp)={}&(\epsilon\cdot q_\perp)
\left[f_1+\frac{\not{P}}{M}f_2+\frac{\not{q}_\perp}{M}f_3
+\frac{\not{P}\not{q}_\perp}{M^2}f_4\right]
+M\not{\epsilon}\,f_5+\not{\epsilon}\not{P}\,f_6 \nonumber\\
&+(\not{q}_\perp\not{\epsilon}-\epsilon\cdot q_\perp)f_7
+\frac{1}{M}\left(\not{P}\not{\epsilon}\not{q}_\perp
-\not{P}\,\epsilon\cdot q_\perp\right)f_8 .
\label{eq:phi1_full}
\end{align}
Only $f_3,f_4,f_5,$ and $f_6$ are independent; $f_1,f_2,f_7,$ and $f_8$ are fixed algebraically by the Salpeter constraints $\varphi^{+-}=\varphi^{-+}=0$, following the conventions of Refs.~\cite{Chang:2003ua,Chang:2004im,Wang:2022cxy}. The $f_5,f_6$ structures are the leading ${}^3S_1$ components, $f_3,f_4$ carry the $D$-wave component, and the constraint-generated terms complete the relativistic $P$-wave sector~\cite{Wang:2022cxy}. Thus the node structure discussed below is a property of the relativistic Salpeter amplitude rather than of a single nonrelativistic radial function.

\subsection{M1 current, form-factor extraction, and width conventions}

Using the Mandelstam representation in the instantaneous approximation~\cite{Mandelstam:1955sd,Ke:2010vx}, the reduced hadronic current is the sum of quark- and antiquark-emission contributions, $J^\mu = J^\mu_b + J^\mu_{\bar c}$, where
\begin{align}
J^\mu_b &= Q_1\int\frac{d^3q}{(2\pi)^3}\, \text{Tr}\left[\frac{\not{P}_i}{M_i}\,\bar\varphi^{++}_{P_f}(\mathbf{q}+\alpha_2\mathbf P_{f\perp})\,\gamma^\mu\,\varphi^{++}_{P_i}(\mathbf{q})\right],\label{eq:Jquark}\\
J^\mu_{\bar c} &= Q_2\int\frac{d^3q}{(2\pi)^3}\, \text{Tr}\left[\bar\varphi^{++}_{P_f}(\mathbf{q}-\alpha_1\mathbf P_{f\perp})\,\frac{\not{P}_i}{M_i}\,\varphi^{++}_{P_i}(\mathbf{q})\,\gamma^\mu\right],\label{eq:Jantiquark}
\end{align}
with $Q_1=Q_b=-1/3$, $Q_2=Q_{\bar c}=-2/3$ (in units of $|e|$), and $\alpha_{1,2}=m_{1,2}/(m_1+m_2)$. The final-state momentum transverse to the initial-state momentum is defined by $P_{f\perp}^{\mu}=P_f^\mu-(P_f\cdot P_i/M_i^2)P_i^\mu$. In the initial-state rest frame, $P_{f\perp}^{\mu}=(0,0,0,-k_\gamma)$, whose spatial component $\mathbf P_{f\perp}$ enters the recoil shifts in Eqs.~\eqref{eq:Jquark} and \eqref{eq:Jantiquark}. The separation $J^\mu=J_b^\mu+J_{\bar c}^\mu$ provides a quark-line-resolved view of the transition amplitude. The two line contributions are not separately observable; only their sum defines the physical transition current.

For $V\to P\gamma$, Lorentz covariance and parity give
\begin{equation}
\langle P(P_f)|J^\mu_{\rm em}|V(P_i,\epsilon)\rangle
=iF_{VP}(Q^2)\epsilon^{\mu\nu\rho\sigma}\epsilon_\nu P_{i\rho}P_{f\sigma}.
\label{eq:lorentz_F}
\end{equation}
Here $Q^\mu=P_i^\mu-P_f^\mu$ is the four-momentum transferred to the emitted photon, with $Q^2=0$ for a real photon.
To make the extraction of $F(0)$ explicit, we work in the initial-state rest frame,
\begin{equation}
P_i^\mu=(M_i,0,0,0),\qquad P_f^\mu=(E_f,0,0,-k_\gamma),
\end{equation}
and choose a transverse vector polarization $\epsilon^\mu=(0,1,0,0)$. Equation~\eqref{eq:lorentz_F} then gives, up to the overall phase fixed by the Levi-Civita convention,
\begin{equation}
J^y=\pm iM_i k_\gamma F_{VP}(0),
\end{equation}
so that the convention-independent magnitude used numerically is
\begin{equation}
|F(0)| = \frac{|J^y|}{M_i k_\gamma},\qquad k_\gamma=\frac{M_i^2-M_f^2}{2M_i}.
\label{eq:F_extract}
\end{equation}
The same construction applies to $P\to V\gamma$ with the transverse polarization assigned to the final vector state. The corresponding radiative widths are
\begin{align}
\Gamma[V\to P\gamma] &= \frac{\alpha_{\rm em}}{3}\,k_\gamma^3\,|F_{VP}(0)|^2,\\
\Gamma[P\to V\gamma] &= \alpha_{\rm em}\,k_\gamma^3\,|F_{PV}(0)|^2,
\label{eq:PtoVwidth}
\end{align}
where $\alpha_{\rm em}=1/137.036$. The different prefactors arise from
the average over the spin states of the initial particle: the initial
vector state carries the factor $1/(2J_i+1)=1/3$, whereas the initial
pseudoscalar has $J_i=0$ and therefore no spin-averaging factor.
The numerical stability of the Salpeter solutions and transition integrals, including the strongly cancelled hindered channels, is documented in Appendix~A.

\section{Mass Calibration and Salpeter Wave Functions}
\label{sec:wavefunctions}

\subsection{Mass calibration and spectrum}

Following the Salpeter analysis of Ref.~\cite{Wang:2022cxy}, the additive constant $V_0$ is calibrated separately in the $0^-$ and $1^-$ sectors to reproduce the measured $B_c(1S)$ and $B_c^*(1S)$ masses, yielding $M_{B_c(1S)}^{\rm BS}=6273$~MeV (PDG: $6274.47$~MeV~\cite{PDG:2024}) and $M_{B_c^*(1S)}^{\rm BS}=6339$~MeV, respectively. All remaining interaction parameters are common to the two sectors, and the fitted values of $V_0$ are then kept fixed for the corresponding radial excitations. The common input parameters used in the present Salpeter solutions are $m_b=4.96$~GeV, $m_c=1.62$~GeV, $\lambda=0.21$~GeV$^2$, $\alpha=0.06$~GeV, $a=e$, and $\Lambda_{\rm QCD}=0.27$~GeV. The sector-dependent values of $V_0$ are given in Appendix~A.

The resulting BS eigenmasses for the $1S$ and $2S$ states are listed in Table~\ref{tab:masses}.
The ground-state masses are within $1.5$~MeV of the experimental values by construction.
The $2S$ masses are predictions of the frozen kernel: $M[B_c(2S)]=6853$~MeV and $M[B_c^*(2S)]=6918$~MeV, which are consistent with the CMS measurement of $M[B_c(2S)]=6871.0$~MeV~\cite{CMS:2019uhm} within $\sim18$~MeV, a typical level of agreement for Salpeter model predictions of excited $B_c$ states~\cite{Ding:2021dwh}.

\begin{table}[h]
\caption{Salpeter eigenmasses and node counts. The $V_0$ values are calibrated to the experimental ground-state masses and frozen for all radial excitations. Experimental masses: $B_c(1S)$ from Ref.~\cite{PDG:2024}; $B_c^*(1S)$ derived from the ATLAS hyperfine splitting of Eq.~\eqref{eq:delta_hf} added to the $B_c(1S)$ mass, with the combined uncertainty ${}^{+1.7}_{-2.0}$~MeV from propagating the statistical and systematic errors in quadrature; $B_c(2S)$ from CMS~\cite{CMS:2019uhm}.}
\label{tab:masses}
\begin{tabular*}{\columnwidth}{@{\extracolsep{\fill}}ccccc@{}}
\hline\hline
State & $J^P$ & $M_{\rm BS}$ (MeV) & $M_{\rm exp}$ (MeV) & Nodes \\
\hline
$B_c(1S)$     & $0^-$ & $6273$ & $6274.47$~\cite{PDG:2024} & 0 \\
$B_c^*(1S)$   & $1^-$ & $6339$ & $6338.97^{+1.7}_{-2.0}$~\cite{ATLAS:2026ubk}  & 0 \\
$B_c(2S)$     & $0^-$ & $6853$ & $6871.0$~\cite{CMS:2019uhm}  & 1 \\
$B_c^*(2S)$   & $1^-$ & $6918$ & ---        & 1 \\
\hline\hline
\end{tabular*}
\end{table}

The frozen-$V_0$ spectrum gives a $2S$ hyperfine splitting of
$65$~MeV. Since the $B_c^*(2S)$ mass has not yet been measured
directly, in Sec.~\ref{sec:results} we also employ experiment-informed
kinematics derived from the observed $2S$ structures and the measured
ground-state hyperfine splitting. This allows the kinematic sensitivity
of the radiative widths to be separated from the bound-state dynamics.

\subsection{Radial wave functions and node structure}

The $0^-$ wave functions are specified by the two independent radial functions $g_1(q)$ and $g_2(q)$, while the $1^-$ wave functions are specified by $f_3(q)$ through $f_6(q)$, as defined in Sec.~\ref{sec:formalism}. Their numerical quality and normalization are summarized in Appendix~A.

The $1S$ wave functions are nodeless in all independent components, consistent with ground-state radial structure. In the $2S$ pseudoscalar state, $g_1$ and $g_2$ cross zero at $q\simeq0.848$~GeV. In the vector $2S$ state, the leading $S$-wave radial functions $f_5$ and $f_6$ cross zero at $q\simeq0.855$~GeV, whereas the $D$-wave radial functions $f_3$ and $f_4$ have zeros near $0.810$ and $0.881$~GeV, respectively. We therefore use $q\simeq0.85$~GeV as the characteristic $2S$ node region, noting that the component-dependent zero positions span approximately $\pm0.04$~GeV around this value.

\begin{figure}[!t]
\centering
\includegraphics[width=0.9\columnwidth]{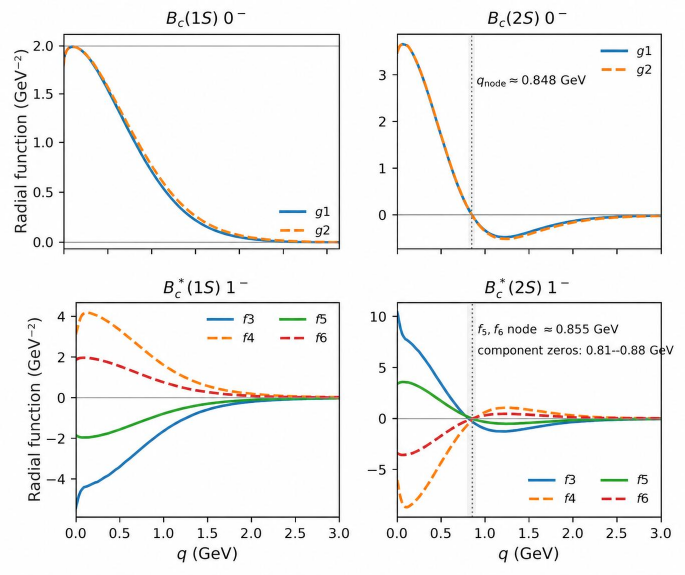}
\caption{Independent radial wave functions $g_1,g_2$ for the $0^-$ states and $f_3,f_4,f_5,f_6$ for the $1^-$ states, shown for the $1S$ and $2S$ levels, in units of GeV$^{-2}$. Here $q=|\mathbf q|=\sqrt{-q_\perp^2}$ denotes the magnitude of the relative three-momentum in the meson rest frame. In the $B_c(1S)$ $0^-$ panel, the overall phase convention is chosen so that the displayed pseudoscalar radial functions $g_1$ and $g_2$ are positive; the other panels retain the original sign convention used in the calculation. The $2S$ pseudoscalar functions $g_1,g_2$ have a common zero near $0.848$~GeV; in the vector $2S$ state the leading $S$-wave functions $f_5,f_6$ cross near $0.855$~GeV, while the $D$-wave functions $f_3,f_4$ have component-dependent zeros near $0.810$ and $0.881$~GeV. The shaded band marks the characteristic node region used in the discussion.}
\label{fig:wavefunctions}
\end{figure}

The independent radial wave functions are shown in Fig.~\ref{fig:wavefunctions}.

\section{M1 Form Factors and Widths}
\label{sec:results}

We consider the four M1 channels
\begin{align}
\text{C1:}&\quad B_c^*(1S)\to B_c(1S)\gamma,\\
\text{C2:}&\quad B_c^*(2S)\to B_c(2S)\gamma,\\
\text{C3:}&\quad B_c^*(2S)\to B_c(1S)\gamma,\\
\text{C4:}&\quad B_c(2S)\to B_c^*(1S)\gamma.
\end{align}
C1 and C2 are allowed transitions with unchanged radial quantum number, whereas C3 and C4 are hindered $2S\to1S$ transitions. The hindered channels involve a change of radial quantum number, so their leading overlaps are expected to be particularly sensitive to radial-node cancellation.

For each channel, we present results under two kinematic schemes:
\begin{itemize}
\item \textbf{Scheme M} (model-consistent): $M_{\rm external}=M_{\rm BS}$, using the Salpeter eigenmasses.
\item \textbf{Scheme E} (experiment-informed): $M_{\rm external}$ uses experimental masses where available. For $B_c(1S)$ and $B_c^*(1S)$, we use the PDG and ATLAS values, respectively. For $B_c(2S)$, we use the CMS measurement of $6871.0$~MeV. For $B_c^*(2S)$, which has not been directly measured, we use the standard CMS interpretation of the two reconstructed $2S$ structures: the higher-mass peak is assigned to $B_c(2S)$, while the lower reconstructed peak is consistent with $B_c^*(2S)\to B_c^*(1S)\pi\pi$ followed by the undetected soft photon from $B_c^*(1S)\to B_c(1S)\gamma$~\cite{CMS:2019uhm,LHCb:2019bem}. The measured peak separation therefore satisfies $\Delta_{\rm peak}=\Delta_{\rm HF}(1S)-\Delta_{\rm HF}(2S)$. Using $\Delta_{\rm peak}=29.1$~MeV and the ATLAS value $\Delta_{\rm HF}(1S)=64.5$~MeV gives $\Delta_{\rm HF}(2S)=35.4$~MeV and hence $M[B_c^*(2S)]=6871.0+35.4=6906.4$~MeV. This is an experiment-informed inference under the standard peak assignment, not a direct measurement of $M[B_c^*(2S)]$.
\end{itemize}

The complete results are presented in Table~\ref{tab:results}.
The numerical transversality of the electromagnetic current has been verified for all channels; the corresponding checks are summarized in Appendix~\ref{sec:app_stability}.

\begin{table}[t]
\caption{M1 form factors and radiative widths. ``Sch.'' denotes kinematic scheme: Sch.~M = model-consistent (Salpeter eigenmasses); Sch.~E = experiment-informed. For C1 and C2 under Scheme~E, the quoted width uncertainties are experimental-kinematic (from propagation of the external mass inputs; see Sec.~\ref{sec:phasespace}). The form-factor uncertainties estimated from the convergence study in Appendix~A are $0.1\%$ for C1 and $2\%$ for C3 and C4; the corresponding numerical uncertainties in the widths are approximately twice as large because $\Gamma\propto |F(0)|^2$. For C3 and C4 under Scheme~E the kinematic sensitivity is small (Table~\ref{tab:phasespace}); only numerical uncertainties are quoted.}
\label{tab:results}
\begin{tabular*}{\columnwidth}{@{\extracolsep{\fill}}ccccc@{}}
\hline\hline
Channel & Sch. & $|F(0)|$ & $k_\gamma$ & $\Gamma$ \\
 & & (GeV$^{-1}$) & (GeV) & (keV) \\
\hline
C1 & E & $0.2854$ & $0.0642$ & $(5.24^{+0.43}_{-0.47})\times10^{-2}$ \\
C1 & M & $0.2854$ & $0.0657$ & $5.61\times10^{-2}$ \\
C2 & E & $0.2759$ & $0.0353$ & $(8.15^{+2.1}_{-1.9})\times10^{-3}$ \\
C2 & M & $0.2746$ & $0.0647$ & $4.97\times10^{-2}$ \\
C3 & E & $0.0105$ & $0.6030$ & $(5.85\pm0.23)\times10^{-2}$ \\
C3 & M & $0.0094$ & $0.6149$ & $5.04\times10^{-2}$ \\
C4 & E & $0.0120$ & $0.5114$ & $(1.40\pm0.06)\times10^{-1}$ \\
C4 & M & $0.0133$ & $0.4947$ & $1.55\times10^{-1}$ \\
\hline\hline
\end{tabular*}
\end{table}

The allowed transitions C1 and C2 have similar form factors. For the hindered transitions C3 and C4, the form factors are smaller than that of C1 by factors of about 27 and 24, respectively. These ratios compare different transitions and should not be confused with the radial cancellation factor $\mathcal C$, defined in Sec.~\ref{sec:hierarchy} as the ratio between the integrated magnitude of the radial contributions and the magnitude of their net integral within a given transition. The latter is $\mathcal C=1$ for C1 and C2 and approximately 18--20 for C3 and C4.

The numerical stability of representative allowed and hindered amplitudes has been checked through convergence tests. We conservatively assign form-factor uncertainties of $0.1\%$ to C1 and $2\%$ to C3 and C4; the corresponding numerical width uncertainties are approximately $0.2\%$ and $4\%$, respectively. The supporting convergence results are given in Appendix~A.

It is important to distinguish the suppression of the hindered transition form factors from that of the radiative widths. Relative to the ground-state allowed channel C1, the form factors of C3 and C4 are smaller by factors of approximately $27$ and $24$, respectively. However, the corresponding photon-momentum factors $k_\gamma^3$ are larger by approximately $8.3\times10^2$ and $5.1\times10^2$. As a result, the C3 width is comparable to the C1 width, with $\Gamma_{C3}/\Gamma_{C1}\simeq1.12$. The C4 width is approximately $2.7$ times the C1 width; in this comparison, the $P\to V\gamma$ and $V\to P\gamma$ widths also differ by a factor of three due to the initial-spin averaging factors in Eqs.~\eqref{eq:PtoVwidth}. Thus, radial-node cancellation strongly suppresses the hindered form factors, but the larger photon momenta compensate for this effect in the physical radiative widths.

\section{Kinematics versus Bound-State Dynamics}
\label{sec:phasespace}

The radiative width can be separated into a kinematic factor and a bound-state factor:
\begin{equation}
\Gamma \propto \underbrace{k_\gamma^3}_{\text{kinematic factor}} \times \underbrace{|F(0)|^2}_{\text{bound-state dynamics}}.
\end{equation}
Before the ATLAS measurement, the $B_c^*$ mass was not directly known, and theoretical predictions for $\Gamma[B_c^*\to B_c\gamma]$ combined uncertainties from both the photon momentum and the form factor.
With the experimental determination of $M[B_c^*(1S)]$, the kinematic factor is fixed by experiment, and the remaining model-dependent uncertainty in the ground-state width is primarily associated with $|F(0)|^2$.

\subsection{Kinematic and dynamical regimes}
\label{sec:regimes}

The four channels fall into three distinct regimes of kinematic sensitivity:

\noindent\textbf{C1: Kinematics fixed by experiment.}
For the ground-state transition, the ATLAS measurement reduces the $k_\gamma$ uncertainty from the model spread of approximately $55$--$66$~MeV in hyperfine splitting to a measured value of $64.5$~MeV.
The $k_\gamma^3$ ratio between Scheme~M and Scheme~E is $0.934$, and the form factor is unchanged ($|F(0)|=0.2854$~GeV$^{-1}$ in both schemes).
The C1 width under experiment-informed kinematics is $\Gamma=5.24\times10^{-2}$~keV, which constitutes our prediction for the $B_c^*(1S)\to B_c(1S)\gamma$ radiative width.
Propagating the experimental uncertainty of $\Delta_{\rm HF}(1S)=64.5^{+1.7}_{-2.0}$~MeV (after combining the statistical and systematic uncertainties in quadrature) while keeping the form factor fixed gives
\begin{equation}
\Gamma_{\rm C1}=\left(5.24^{+0.43}_{-0.47}\right)\times10^{-2}~\text{keV},
\end{equation}
corresponding to an experimental-kinematic uncertainty of $+8.2\%/-8.9\%$.
Since this radiative transition is expected to dominate the total width of the $B_c^*$ meson below the open-flavor threshold, the predicted radiative width also provides an estimate of the $B_c^*$ natural width.

\noindent\textbf{C2: Kinematic-factor dominated.}
The $B_c^*(2S)$ mass has not been directly measured, and its inferred value depends on the unknown $2S$ hyperfine splitting.
The photon momentum changes from $k_\gamma=0.0647$~GeV (Scheme~M) to $k_\gamma=0.0353$~GeV (Scheme~E, derived from the CMS $2S$ peak relation and ATLAS $B_c^*(1S)$ mass), a factor of $1.83$.
The $k_\gamma^3$ ratio is $0.163$, and the width changes by a factor of approximately $6.1$ purely from the kinematic factor.
The form factor $|F_{22}(0)|$ remains stable at $0.275$--$0.276$~GeV$^{-1}$.
Thus the present C2 width is controlled mainly by the unknown $B_c^*(2S)$ mass, and a direct measurement of this state would largely remove this uncertainty.
A direct observation of the soft photon in $B_c^*(2S)\to B_c(2S)\gamma$, whose energy directly probes the $2S$ hyperfine splitting, would also test the assignment of the structures observed in the $2S$ region.
The derived Scheme-E width carries an experimental-kinematic uncertainty of approximately $+26\%/-23\%$, reflecting the experimental uncertainties of the CMS $B_c(2S)$ mass, the CMS peak separation, and the ATLAS ground-state hyperfine splitting propagated through the inferred $B_c^*(2S)$ mass. The quoted uncertainty is obtained by combining the experimental input uncertainties in quadrature and neglecting correlations among them, since no covariance information is used in the present analysis. A direct measurement of $B_c^*(2S)$ would remove this remaining kinematic ambiguity.

\noindent\textbf{C3 and C4: Dynamics dominated.}
For the hindered $2S\to1S$ transitions, the photon momentum is large ($k_\gamma\approx0.5$--$0.6$~GeV) and nearly identical between the two kinematic schemes.
The widths are therefore robust to the remaining kinematic uncertainties, and the form-factor suppression by a factor of $\sim25$ relative to the ground-state allowed channel C1 is a genuine dynamical effect driven by radial-node cancellation. Here, ``dynamics dominated'' refers to the relative stability under changes of the external mass inputs between Schemes M and E, rather than to a suppression of the hindered radiative widths relative to C1.

Table~\ref{tab:phasespace} summarizes the kinematics/dynamics decomposition.

\begin{table}[t]
\caption{Kinematics versus dynamics decomposition. All ratios are defined as Scheme~E divided by Scheme~M.}
\label{tab:phasespace}
\begin{tabular*}{\columnwidth}{@{\extracolsep{\fill}}cccc@{}}
\hline\hline
Channel & $k_\gamma^3$ ratio & $|F(0)|^2$ ratio & $\Gamma$ ratio \\
\hline
C1 & $0.934$ & $1.000$ & $0.934$ \\
C2 & $0.163$ & $1.010$ & $0.164$ \\
C3 & $0.943$ & $1.233$ & $1.162$ \\
C4 & $1.105$ & $0.818$ & $0.904$ \\
\hline\hline
\end{tabular*}
\end{table}

The small $\Gamma$ ratio for C2 does not originate from a change in the form factor; $|F(0)|^2$ differs by only $1\%$ between the two schemes.  Instead, the suppression is almost entirely kinematic: the photon momentum decreases from $0.0647$~GeV (Scheme~M) to $0.0353$~GeV (Scheme~E), and the $k_\gamma^3$ dependence therefore suppresses the width by a factor of approximately six, while $|F_{22}(0)|$ remains essentially unchanged.

\subsection{Comparison with other theoretical predictions}

Table~\ref{tab:literature} compares our M1 predictions with representative results from the Bethe--Salpeter approach~\cite{Ke:2010vx}, nonrelativistic quark-potential models~\cite{Li:2019qkt,Asghar:2019abc,Gao:2024yvz}, a constituent-quark model~\cite{MartinGonzalez:2022bc}, the light-front quark model~\cite{Ridwan:2024ngc}, and a screened-potential model~\cite{Patel:2026prg}. The quoted widths are those reported in the respective publications, converted to keV where necessary; they have not been recalculated using a common set of masses. Since M1 widths depend strongly on $k_\gamma^3$, differences among the published values reflect both the dynamical treatment and the chosen mass splittings. The form factors are listed separately where available.

\begin{table}[t]
\caption{Comparison with representative theoretical predictions for the four $B_c$ M1 transitions C1--C4. Radiative widths $\Gamma$ are given in keV and transition form factors $|F(0)|$ in GeV$^{-1}$. The values from the literature are quoted as published, with eV converted to keV where necessary; no common-mass rescaling is applied. Different choices of meson masses and radiative conventions affect the widths, particularly for the allowed transitions. Particular care is needed when comparing the C4 widths, because the spin-averaging conventions for $P\to V\gamma$ and $V\to P\gamma$ should be checked across different calculations. A dash denotes a value not reported in the cited work. Uncertainties on our predictions are given in Table~\ref{tab:results}.}
\label{tab:literature}
\footnotesize
\setlength{\tabcolsep}{4pt}
\begin{tabular*}{\columnwidth}{@{\extracolsep{\fill}}lcccc@{}}
\hline\hline
Model & C1 & C2 & C3 & C4 \\
\hline
\multicolumn{5}{c}{$\Gamma$ (keV)} \\
\hline
Ke et al.~\cite{Ke:2010vx}                    & $0.017$ & --- & $0.28$ & $0.38$ \\
Li et al.~\cite{Li:2019qkt}                   & $0.057$ & $0.0024$ & $1.205$ & $0.099$ \\
Asghar et al.~\cite{Asghar:2019abc}           & $0.15$  & $1.4\times10^{-4}$ & $0.41$ & $0.07$ \\
Mart\'in-Gonz\'alez et al.~\cite{MartinGonzalez:2022bc} & $0.052$ & $0.010$ & $0.650$ & $0.250$ \\
Gao et al.~\cite{Gao:2024yvz}                 & $0.040$ & $0.0033$ & $0.56$ & $0.14$ \\
Ridwan et al.~\cite{Ridwan:2024ngc}           & $0.0040$ & $0.0015$ & $0.57$ & $1.38$ \\
Patel et al.~\cite{Patel:2026prg}             & $0.063$ & $0.0047$ & $0.32$ & $9.92$ \\
This work                                    & $0.052$ & $0.0082$ & $0.059$ & $0.140$ \\
\hline
\multicolumn{5}{c}{$|F(0)|$ (GeV$^{-1}$), where available} \\
\hline
Ke et al.~\cite{Ke:2010vx}                    & $0.208$ & --- & $0.023$ & $0.030$ \\
Ridwan et al.~\cite{Ridwan:2024ngc}           & $0.299$ & $0.281$ & $0.034$ & $0.034$ \\
This work                                    & $0.285$ & $0.276$ & $0.0105$ & $0.0120$ \\
\hline\hline
\end{tabular*}
\end{table}

The expanded comparison shows a substantial spread among the published hindered-transition widths. Part of this spread arises from the different input masses and photon momenta, while the remaining differences reflect the treatment of the transition matrix elements. In the present calculation, the radial and quark-line decompositions below are used to examine the cancellation mechanism directly at the form-factor level, without attributing the full spread of published radiative widths to radial dynamics alone.

\section{Radial-Node Filtering in Allowed and Hindered Transitions}
\label{sec:hierarchy}

\subsection{Radial integrand and cumulative form factor}

To quantitatively characterize the node-induced cancellation in the hindered M1 transitions, we define a radial form-factor integrand $I_F(q)$ such that
\begin{equation}
F(0) = \int_0^\infty dq\, I_F(q).
\end{equation}
Explicitly, $I_F(q)$ is obtained from the transverse current component $J^y$ in Eqs.~\eqref{eq:Jquark}--\eqref{eq:Jantiquark} after performing the angular integrations over $\cos\theta$ and $\phi$, including the radial Jacobian $q^2$, the recoil-shifted final-state Salpeter amplitudes $\bar\varphi_{P_f}^{++}(\mathbf{q}\pm\alpha_{1,2}\mathbf P_{f\perp})$, and the overall conversion factor $1/(M_i k_\gamma)$ from Eq.~\eqref{eq:F_extract}. Both quark-emission ($b$) and antiquark-emission ($\bar c$) current contributions are included, so that the integrand decomposes as
\begin{equation}
I_F(q)=I_F^{(b)}(q)+I_F^{(\bar c)}(q).
\end{equation}
The cumulative form factor is
\begin{equation}
F_{\rm cum}(q_c) = \int_0^{q_c} dq\, I_F(q),
\end{equation}
which approaches the signed full form factor at the upper integration boundary $q_c=q_{\max}$.

\begin{figure}[t]
\centering
\includegraphics[width=\columnwidth]{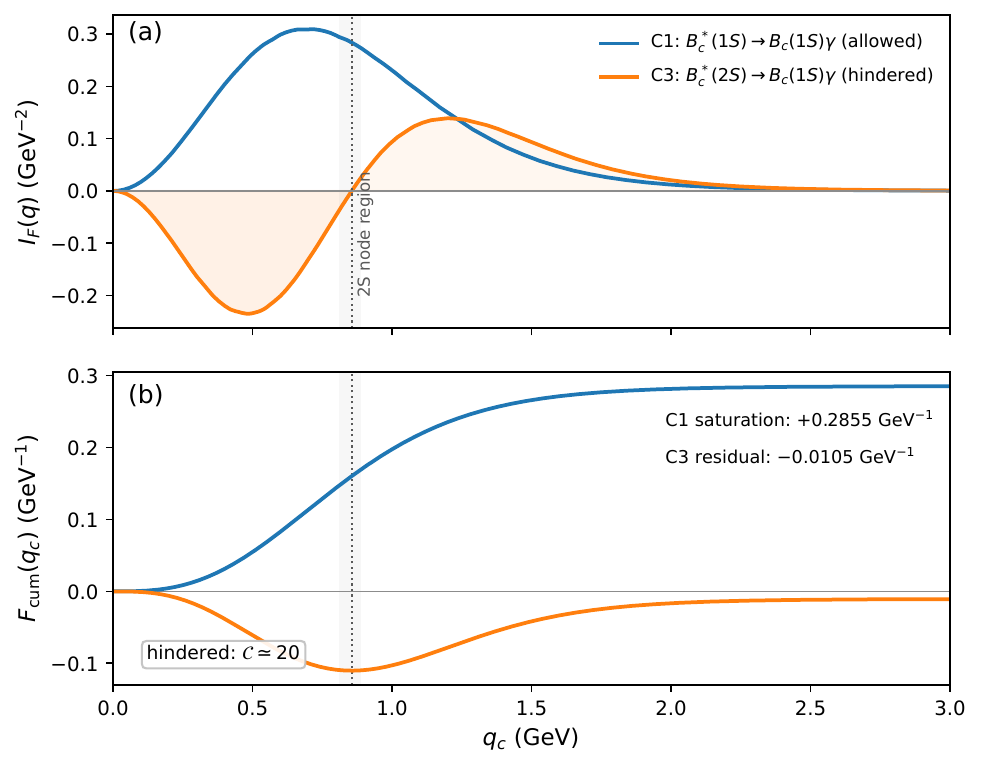}
\caption{(a) Radial form-factor integrand $I_F(q)$ and (b) cumulative form factor $F_{\rm cum}(q_c)$ for the allowed C1 $(1S\to1S)$ and hindered C3 $(2S\to1S)$ transitions.
The cumulative form factor approaches the signed full form factor at $q_c=q_{\max}$.
The hindered channel exhibits a clear sign change at the $2S$ node position ($q\approx0.85$~GeV) and strong cancellation in the cumulative integral, while the allowed channel accumulates smoothly to its saturation value.}
\label{fig:integrand}
\end{figure}

Figure~\ref{fig:integrand} shows $I_F(q)$ for the allowed C1 and the hindered C3 transitions.
The allowed channel exhibits a smooth, sign-definite integrand that accumulates steadily to its final value.
In contrast, the hindered channel shows a clear sign change at the $2S$ node position, with positive and negative regions of comparable magnitude, leading to a strong cancellation in the cumulative integral.

\subsection{Cancellation factor}

We define the cancellation factor
\begin{equation}
\mathcal C = \frac{\displaystyle\int dq\,|I_F(q)|}{\displaystyle\left|\int dq\,I_F(q)\right|}.
\label{eq:cancel}
\end{equation}
The numerator adds the magnitudes of the contributions from different radial-momentum regions, whereas the denominator is the magnitude of the net form factor after contributions of opposite signs have been combined. Thus $1/\mathcal C$ gives the fraction of the integrated absolute contribution that remains in the net integral. A value $\mathcal C\simeq20$ means that the net result is about $5\%$ of the integrated absolute contribution. This quantity measures cancellation within one transition; it is different from the form-factor ratio between hindered and allowed channels.

The computed values are:
\begin{equation}
\mathcal C_{\rm C1}=1.0,\;\mathcal C_{\rm C2}=1.0,\;\mathcal C_{\rm C3}=20.0,\;\mathcal C_{\rm C4}=18.1.
\end{equation}
For C1 and C2, $\mathcal C=1$ indicates the absence of cancellation between radial regions of opposite signs. For C3 and C4, $\mathcal C=20.0$ and $18.1$ imply that about $5.0\%$ and $5.5\%$, respectively, of the integrated absolute radial contributions remain in the net form factors (as shown in Table~\ref{tab:app_integrand}).

\subsection{Localized node-region sensitivity}
\label{sec:node_region}

The cancellation factor identifies the hindered amplitudes as small residuals of positive and negative radial contributions. To quantify how the hindered amplitudes respond to the radial structure around the node, we apply the localized sensitivity test described in Appendix~\ref{sec:app_node_shift}. A local radial-coordinate remapping is applied to all independent components of the $2S$ Salpeter amplitudes while keeping the $1S$ wave functions unchanged, with the target zero defined by the leading $S$-wave components: $f_5,f_6$ for the vector state and $g_1,g_2$ for the pseudoscalar. The remap is confined to a $\pm0.25$~GeV window around the original node region. Moving the leading-$S$ zero from its original value near $0.85$~GeV to $0.70$~GeV changes $|F_{\rm C3}|$ from $0.0105$ to $0.0175$~GeV$^{-1}$ and $|F_{\rm C4}|$ from $0.0120$ to $0.0163$~GeV$^{-1}$; moving it to $1.00$~GeV gives $0.0390$ and $0.0432$~GeV$^{-1}$, respectively. The response is therefore strongly asymmetric in the direction of the remap, and in both channels it confirms substantial sensitivity to the radial structure in the node region.

The strong asymmetric response shows that the small hindered amplitudes originate from a delicate cancellation controlled by the radial structure in the node region. This behavior contrasts with the sign-definite radial integrands of the allowed transitions, for which $\mathcal C=1$.

\subsection{Quark-line-resolved node filtering}
\label{sec:quarkline_main}

We now examine separately the two terms in the transition current: photon emission from the $b$ quark and from the $\bar c$ antiquark. As shown in Appendix~\ref{sec:app_amplitude}, their contributions interfere destructively. For C1--C3, $|J_b|/|J_{\bar c}|\simeq0.18$--$0.20$, while for C4 the ratio is about $0.02$. To determine why the net $b$ contribution is particularly small in C4, we resolve each contribution into its radial-momentum integrand,
\begin{equation}
I_F(q)=I_F^{(b)}(q)+I_F^{(\bar c)}(q),
\end{equation}
and define the line-resolved cancellation factors
\begin{equation}
\mathcal C_b=\frac{\int dq\,|I_F^{(b)}(q)|}{\left|\int dq\,I_F^{(b)}(q)\right|},\qquad
\mathcal C_{\bar c}=\frac{\int dq\,|I_F^{(\bar c)}(q)|}{\left|\int dq\,I_F^{(\bar c)}(q)\right|}.
\end{equation}
Because the electric charge multiplies each line density uniformly, it cancels identically in $\mathcal C_i$. The cancellation hierarchy is therefore not a trivial consequence of $|Q_b|<|Q_{\bar c}|$. For each emission term, $1/\mathcal C_i$ measures the magnitude of its net radial integral relative to the integral of the absolute value of its radial contributions.

For the hindered channels the direct three-dimensional decomposition gives
\begin{equation}
\begin{array}{l|ccc|cc}
 & F_b & F_{\bar c} & F_{\rm tot} & \mathcal C_b & \mathcal C_{\bar c}\\
\hline
{\rm C3} & +0.00265 & -0.01313 & -0.01048 & 20.2 & 20.0\\
{\rm C4} & +0.00025 & -0.01224 & -0.01199 & 2.2\times10^{2} & 22.3
\end{array}
\end{equation}
(all form factors in GeV$^{-1}$). For C3, the cancellation factors of the $b$ and $\bar c$ contributions are both about 20, so roughly $5\%$ of each integrated absolute radial contribution remains in its net form factor. In C4, however, the cancellation factor is about 220 for the $b$ term and about 22 for the $\bar c$ term. Thus only about $0.46\%$ of the $b$ contribution remains after radial integration, compared with about $4.5\%$ for the $\bar c$ contribution. The corresponding ratios after dividing out the constituent charges are $|J_b/Q_b|/|J_{\bar c}/Q_{\bar c}|\simeq0.40$ for C3 and $\simeq0.04$ for C4.

To distinguish a small net form factor from small contributions at individual radial momenta, we introduce the integrated absolute radial contribution for each photon-emission term,
\begin{equation}
U_i\equiv\int dq\,|I_F^{(i)}(q)|=\mathcal C_i|F_i|.
\end{equation}
Using a dense radial reconstruction ($\Delta q=0.01$~GeV), we find
\begin{equation}
\begin{array}{c|cc}
 & U_b & U_{\bar c}\\
\hline
{\rm C3} & 0.0535 & 0.264\\
{\rm C4} & 0.0552 & 0.273
\end{array}\qquad {\rm GeV}^{-1}.
\end{equation}
The integrated absolute $b$ contributions are nearly the same in C3 and C4: $U_b=0.0535$ and $0.0552$~GeV$^{-1}$, respectively. Their net contributions are very different, however: $|F_b|=0.00265$ and $0.00025$~GeV$^{-1}$. Thus the small net $b$ contribution in C4 results from more complete cancellation between radial regions of opposite signs, not from a smaller contribution before cancellation. The quantities $U_i$ characterize the radial integrands within this calculation and are not separately observable widths.

\begin{figure}[t]
\centering
\includegraphics[width=\columnwidth]{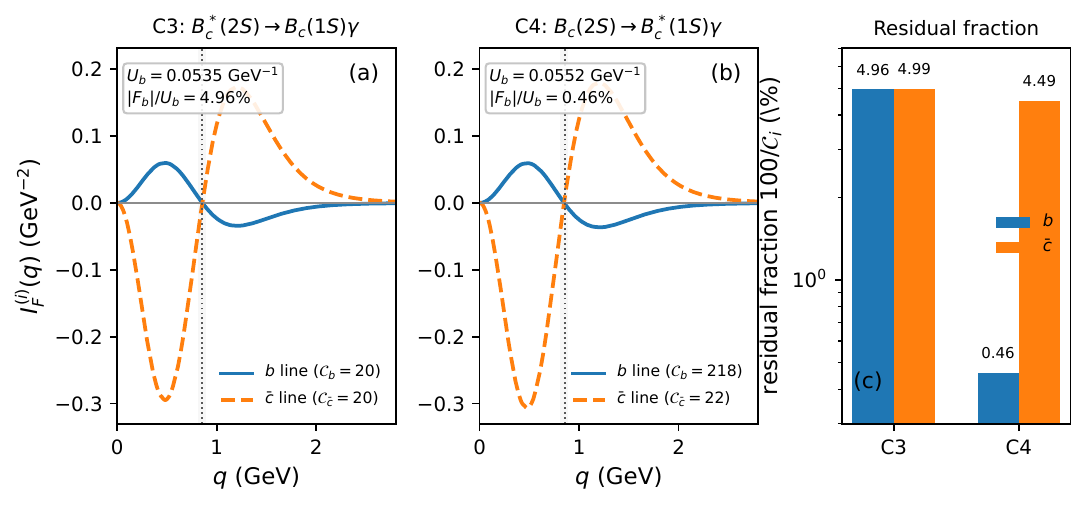}
\caption{Radial contributions from photon emission by the $b$ quark and $\bar c$ antiquark in the hindered C3 and C4 transitions. Panels (a) and (b) show the corresponding radial form-factor integrands $I_F^{(i)}(q)$. The quantity $U_i=\int dq\,|I_F^{(i)}(q)|$ measures the integrated absolute contribution before cancellations between regions of opposite signs, while $|F_i|/U_i=1/\mathcal C_i$ is the fraction remaining in the net integral. Panel (c) compares these fractions (in percent) on a logarithmic scale. Although $U_b$ is nearly the same in C3 and C4, the fraction remaining in the $b$ contribution is about $5\%$ in C3 and only $0.46\%$ in C4.}
\label{fig:quarkline}
\end{figure}

Figure~\ref{fig:quarkline} displays the corresponding continuous radial densities. The line-resolved cancellation factors are stable at the displayed precision under variation of the angular quadrature, with the dense-$q$ reconstruction giving $\mathcal C_b\simeq20.18$ and $\mathcal C_{\bar c}\simeq20.03$ for C3 and $\mathcal C_b\simeq218.3$ and $\mathcal C_{\bar c}\simeq22.3$ for C4. The per-line closure $F_b+F_{\bar c}=F_{\rm total}$ holds at machine precision ($<10^{-14}$), and the reconstructed totals differ from the direct three-dimensional quadrature values by $0.30\%$ (C3) and $0.17\%$ (C4), well within the assigned numerical uncertainty.

A routing test in Appendix~\ref{sec:app_alpha_swap} replaces the physical unequal routing by $\alpha_1=\alpha_2=0.5$ while keeping the masses and Salpeter wave functions fixed. Under this symmetrized routing, $|J_b|/|J_{\bar c}|$ changes from $0.202$ to $0.065$ in C3 and from $0.020$ to $0.045$ in C4. The contrast between the crossed channels is therefore reduced from about a factor of ten to about $1.4$, showing that unequal recoil routing accounts for much of their relative difference. Even with symmetrized routing the $b$ line remains strongly subleading in C4, so the persistent $\bar c$ dominance also reflects the remaining differences between the crossed matrix elements, including the pseudoscalar/vector Salpeter structures and their initial/final-state roles.

\section{Summary}
\label{sec:summary}

We have studied four M1 radiative transitions in the $B_c$ system using relativistic Salpeter wave functions calibrated to the measured ground-state masses. The ATLAS observation of the $B_c^*(1S)$ meson removes the dominant kinematic ambiguity in $B_c^*(1S)\to B_c(1S)\gamma$; using the measured masses gives $|F_{11}(0)|=0.2854$~GeV$^{-1}$ and $\Gamma=5.24\times10^{-2}$~keV. For the allowed $2S\to2S$ transition the form factor remains stable at $|F_{22}(0)|\simeq0.276$~GeV$^{-1}$, but the width is limited by the unmeasured $B_c^*(2S)$ mass. Under the standard CMS peak assignment, the inferred $2S$ hyperfine splitting is $35.4$~MeV, whereas the frozen-$V_0$ Salpeter spectrum gives $65$~MeV; the $k_\gamma^3$ dependence of the radiative width makes C2 particularly sensitive to this kinematic difference.

The allowed transitions exhibit no cancellation between radial regions of opposite signs, with $\mathcal C=1$. For the hindered $2S\to1S$ transitions, the cancellation factors are about 18--20, meaning that only about $5\%$ of the integrated absolute radial contributions remain in the net form factors. The localized node-region test confirms that these form factors are sensitive to the radial structure near the $2S$ node.

Separating photon emission from the $b$ and $\bar c$ constituents reveals an additional difference between the two hindered channels. In C3, both emission terms have cancellation factors of about 20. In C4, the factor increases to about 220 for the $b$ term but remains near 22 for the $\bar c$ term. The integrated absolute $b$ contributions are similar in C3 and C4, whereas the net $b$ contribution is much smaller in C4. The symmetrized-recoil test indicates that the unequal recoil routing is an important source of this difference. Thus, radial-node cancellation controls the hindered M1 form factors, with an additional strong cancellation in the $b$ contribution to C4. The corresponding radiative widths are nevertheless comparable to or larger than the ground-state allowed width because of the much larger photon momenta.

\FloatBarrier
\begin{acknowledgments}
This work was supported by the National Natural Science Foundation of China (NSFC) under Grants No. 12575106 and 12147214, and by the Liaoning Province Fundamental Research Fund under Grant No. LJ212410165019.
\end{acknowledgments}

\paragraph*{AI-assisted tools.} AI-assisted tools---OpenAI ChatGPT (GPT-5.6), DeepSeek V4, GLM 5.2, and Aether (v0.72)---were used for limited code generation and debugging, plotting, reproducibility checks, manuscript revision, and scientific consistency checks. The author specified the physical assumptions, conventions, validation benchmarks, and final scientific judgments; independently checked generated or modified code against those benchmarks; and verified the final numerical results and manuscript content.

\appendix

\section{Numerical details and convergence checks}
\label{sec:app_numerical}

\subsection{Model parameters and Salpeter solutions}

The $0^-$ and $1^-$ Salpeter wave functions are obtained from the complete instantaneous equations described in Sec.~\ref{sec:formalism}. The $V_0^{(0^-)}$ parameter is tuned to reproduce the measured $B_c(1S)$ mass within the numerical calibration accuracy, yielding $M_{B_c(1S)}^{\rm BS}=6273$~MeV; $V_0^{(1^-)}$ is tuned to the ATLAS $B_c^*(1S)$ mass, yielding $M_{B_c^*(1S)}^{\rm BS}=6339$~MeV.
The fitted additive constants used in the present calculation are $V_0^{(0^-)}=-0.189$~GeV and $V_0^{(1^-)}=-0.114$~GeV in the sign convention of Eq.~\eqref{eq:kernel}. Equivalently, the numerical solver uses the positive parameters $C=-V_0=0.189$ and $0.114$~GeV in the two sectors. Both values are frozen for all higher radial excitations.
The common input parameters used to generate the present Salpeter wave functions are $m_b=4.96$~GeV, $m_c=1.62$~GeV, $\lambda=0.21$~GeV$^2$ (linear confinement strength), $\alpha=0.06$~GeV (infrared screening parameter), $a=e=2.71828$, and $\Lambda_{\rm QCD}=0.27$~GeV (QCD scale). Together with the sector-dependent $V_0$ values specified above, these are the parameters of the numerical Salpeter solutions used throughout this work.

The $0^-$ states are ${}^1S_0$ dominant with a small $P$-wave admixture; the $1^-$ states are ${}^3S_1$ dominant with small $P$ and $D$ wave components~\cite{Wang:2022cxy}.

\begin{table}[t]
\caption{Salpeter eigenmasses, numerical residuals, and node counts. Norm.\ denotes the numerical Salpeter normalization integral. For the $2S$ states, ``1'' denotes one radial zero in each independent component; the component-dependent zero positions are discussed in Sec.~\ref{sec:wavefunctions}.}
\label{tab:app_wavefunction}
\begin{tabular*}{\columnwidth}{@{\extracolsep{\fill}}cccccc@{}}
\hline\hline
State & $J^P$ & $M_{\rm BS}$ (MeV) & Residual & Norm. & Nodes \\
\hline
$B_c(1S)$     & $0^-$ & $6273$ & $<10^{-13}$ & $\sim1.00$ & 0 \\
$B_c^*(1S)$   & $1^-$ & $6339$ & $<10^{-13}$ & $\sim1.00$ & 0 \\
$B_c(2S)$     & $0^-$ & $6853$ & $<10^{-13}$ & $\sim1.00$ & 1 \\
$B_c^*(2S)$   & $1^-$ & $6918$ & $<10^{-13}$ & $\sim1.00$ & 1 \\
\hline\hline
\end{tabular*}
\end{table}

\subsection{Numerical solution and convergence}

The complete Salpeter equations of Sec.~\ref{sec:formalism} are solved numerically after discretizing the radial momentum $q\in[0,q_{\rm max}]$ on a grid and evaluating the kernel and Salpeter amplitudes at the grid points. The resulting eigenvalue (bound-state mass) and radial functions satisfy residuals below $10^{-13}$ (Table~\ref{tab:app_wavefunction}). The node positions are extracted as the sign changes of the leading $S$-wave components after convergence. The M1 transition integrals are evaluated on a three-dimensional $(q,\cos\theta,\phi)$ grid using standard quadrature, with $n_q$ radial points, $n_c$ $\cos\theta$ points, and $n_\phi$ azimuthal points. The wave-function support boundary is $q_{\rm max}=3.87$~GeV, beyond which the radial functions are set to zero (see Appendix~\ref{sec:app_stability}).

Table~\ref{tab:app_convergence} summarizes the numerical stability of the M1 form factors under successive increases of the integration resolution.

\begin{table}[t]
\caption{Grid convergence. Here $n_q$, $n_c$, and $n_\phi$ denote the numbers of quadrature points in the radial momentum $q$, the polar angle $\cos\theta$, and the azimuthal angle $\phi$, respectively. $|F(0)|$ is in GeV$^{-1}$ and $\Gamma$ is in keV.}
\label{tab:app_convergence}
\begin{tabular*}{\columnwidth}{@{\extracolsep{\fill}}cccccc@{}}
\hline\hline
Channel & Grid & $(n_q,n_c,n_\phi)$ & $|F(0)|$ & $\Gamma$ & $|\delta F/F|$ \\
\hline
C1 & coarse  & $(36,28,8)$   & $0.285528$ & $5.24\times10^{-2}$ & $0.04\%$ \\
C1 & central & $(48,36,12)$  & $0.285406$ & $5.24\times10^{-2}$ & --- \\
C1 & fine    & $(64,48,16)$  & $0.285530$ & $5.24\times10^{-2}$ & $0.04\%$ \\
C3 & coarse  & $(36,28,8)$   & $0.010682$ & $6.09\times10^{-2}$ & $1.98\%$ \\
C3 & central & $(48,36,12)$  & $0.010474$ & $5.85\times10^{-2}$ & --- \\
C3 & fine    & $(64,48,16)$  & $0.010614$ & $6.01\times10^{-2}$ & $1.33\%$ \\
C4 & coarse  & $(36,28,8)$   & $0.012197$ & $1.45\times10^{-1}$ & $1.70\%$ \\
C4 & central & $(48,36,12)$  & $0.011993$ & $1.40\times10^{-1}$ & --- \\
C4 & fine    & $(64,48,16)$  & $0.012127$ & $1.44\times10^{-1}$ & $1.12\%$ \\
\hline\hline
\end{tabular*}
\end{table}

\subsection{Current transversality and finite-support stability}
\label{sec:app_stability}

Table~\ref{tab:app_gauge} summarizes the transversality and numerical-quality metrics for the four channels under Scheme~E. The Ward-identity residual $Q_\mu J^\mu$ is at machine precision ($10^{-17}$) for all channels, providing a stringent numerical check that the computed current is transverse within the positive-energy Salpeter calculation. The non-M1 current fraction $|J^{\rm non\text{-}M1}|/|J^y|$ (where $J^{\rm non\text{-}M1}$ denotes the current components $J^x$, $J^z$, and $J^0$, which vanish for a pure M1 transition) is at the same level. The outside-support fractions $f_{\rm out}^{(1,2)}$ (defined as the fraction of the integration measure where the recoil-shifted radial argument exceeds the wavefunction support boundary $q_{\rm max}=3.87$~GeV) are below $1\%$ for C1 and C2, and $3$--$9\%$ for the hindered channels C3 and C4, which have larger photon momenta that shift the radial arguments closer to the support boundary. All radial components of the pseudoscalar ($g_i$) and vector ($f_i$) states satisfy $|u(q_{\max})|/|u|_{\rm peak}\lesssim10^{-3}$ at the support boundary, where $u$ denotes any of these radial components.

\begin{table}[t]
\caption{Current-transversality and numerical quality metrics (Scheme~E).}
\label{tab:app_gauge}
\begin{tabular*}{\columnwidth}{@{\extracolsep{\fill}}ccccc@{}}
\hline\hline
Channel & $|Q_\mu J^\mu|/|k_\gamma J^y|$ & $|J_{\rm non-M1}|/|J^y|$ & $f_{\rm out}^{(1)}$ & $f_{\rm out}^{(2)}$ \\
\hline
C1 & $3.3\times10^{-17}$ & $4.1\times10^{-17}$ & $0.30\%$ & $0.95\%$ \\
C2 & $2.1\times10^{-17}$ & $4.7\times10^{-17}$ & $0.18\%$ & $0.49\%$ \\
C3 & $4.0\times10^{-17}$ & $4.5\times10^{-17}$ & $2.96\%$ & $8.62\%$ \\
C4 & $1.0\times10^{-16}$ & $7.8\times10^{-17}$ & $2.45\%$ & $7.55\%$ \\
\hline\hline
\end{tabular*}
\end{table}

The stability of the hindered form factors against enlargement of the integration domain is summarized in Table~\ref{tab:app_qmax}. The default integration cutoff is $q_{\rm max}=3.87$~GeV, coinciding with the wave-function support boundary; beyond this support the wave function is set to zero.

\begin{table}[t]
\caption{Integration-domain stability for the hindered C3 and C4 form factors (Scheme~E) with fixed wave-function support $q_{\rm wf,max}=3.87$~GeV. The relative change is computed against the $q_{\rm max}=3.87$~GeV reference.}
\label{tab:app_qmax}
\begin{tabular*}{\columnwidth}{@{\extracolsep{\fill}}cccc@{}}
\hline\hline
Channel & $q_{\rm max}$ (GeV) & $|F(0)|$ (GeV$^{-1}$) & $\Delta F/F$ \\
\hline
\multirow{3}{*}{C3} & $3.87$ & $0.01048$ & --- \\
                     & $5.00$ & $0.01046$ & $-0.14\%$ \\
                     & $6.00$ & $0.01035$ & $-1.21\%$ \\
\multirow{3}{*}{C4} & $3.87$ & $0.01199$ & --- \\
                     & $5.00$ & $0.01197$ & $-0.17\%$ \\
                     & $6.00$ & $0.01184$ & $-1.32\%$ \\
\hline\hline
\end{tabular*}
\end{table}

Enlarging the numerical integration domain changes the reported hindered form factors by at most about $1.2$--$1.3\%$. Together with the small pseudoscalar ($g_i$) and vector ($f_i$) radial amplitudes at the support boundary, these results establish the numerical stability of the finite-support treatment used for the transition integrals.

\section{Additional amplitude decompositions}
\label{sec:app_amplitude}

\subsection{Quark-line current decomposition}

Table~\ref{tab:app_current} gives the $b$-quark and $\bar c$-antiquark contributions to the $J^y$ current for all four channels under Scheme~E kinematics, used in the line-resolved discussion of Sec.~\ref{sec:quarkline_main}. The two contributions interfere destructively in all channels, with the $b$-quark term suppressed relative to the $\bar c$ term. The C4 channel exhibits the smallest ratio $|J_b|/|J_{\bar c}|\approx0.02$.

\begin{table}[t]
\caption{Decomposition of $J^y$ into $b$-quark and $\bar c$-antiquark contributions (Scheme~E). Current components are given in GeV.}
\label{tab:app_current}
\begin{tabular*}{\columnwidth}{@{\extracolsep{\fill}}cccccc@{}}
\hline\hline
Channel & $|J_b|$ & $|J_{\bar c}|$ & $|J_{\rm total}|$ & $|J_b|/|J_{\bar c}|$ & Type \\
\hline
C1 & $0.0267$ & $0.1428$ & $0.1162$ & $0.187$ & destructive \\
C2 & $0.0148$ & $0.0821$ & $0.0673$ & $0.181$ & destructive \\
C3 & $0.0110$ & $0.0547$ & $0.0436$ & $0.202$ & destructive \\
C4 & $0.0009$ & $0.0430$ & $0.0421$ & $0.020$ & destructive \\ 
\hline\hline
\end{tabular*}
\end{table}

\subsection{Radial-integrand decomposition and cancellation measures}

Table~\ref{tab:app_integrand} separates the radial form-factor integral into its positive and negative contributions, which determine the cancellation factors discussed in Sec.~\ref{sec:hierarchy}. We define
\begin{equation}
P=\int_{I_F>0}dq\,I_F(q),\qquad
N=-\int_{I_F<0}dq\,I_F(q),
\end{equation}
and the cancellation factor is $\mathcal C=(P+N)/|P-N|$.

\begin{table}[t]
\caption{Integrand analysis. $P=\int_{I_F>0}dq\,I_F(q)$, $N=-\int_{I_F<0}dq\,I_F(q)$, $\mathcal C=(P+N)/|P-N|$. $F_{\rm signed}$, $|F(0)|$, $P$, and $N$ are given in GeV$^{-1}$; $\mathcal C$ and $P/N$ are dimensionless. The sign of $F_{\rm signed}$ is fixed by the relative position of the $2S$ node and the dominant overlap region; the physical width depends on $|F(0)|^2$ and is sign-independent.}
\label{tab:app_integrand}
\begin{tabular*}{\columnwidth}{@{\extracolsep{\fill}}ccccccc@{}}
\hline\hline
Channel & $F_{\rm signed}$ & $|F(0)|$ & $P$ & $N$ & $P/N$ & $\mathcal C$ \\
\hline
C1 & $+0.2854$ & $0.2854$ & $0.2854$ & $0.000$ & --- & $1.0$ \\
C2 & $+0.2759$ & $0.2759$ & $0.2759$ & $0.000$ & --- & $1.0$ \\
C3 & $-0.0105$ & $0.0105$ & $0.0998$ & $0.1103$ & $0.905$ & $20.0$ \\
C4 & $-0.0120$ & $0.0120$ & $0.1031$ & $0.1151$ & $0.896$ & $18.1$ \\
\hline\hline
\end{tabular*}
\end{table}

The numerical values reproduce the allowed/hindered separation shown in Sec.~\ref{sec:hierarchy}: the allowed channels have $\mathcal C=1$ (sign-definite integrands), while the hindered channels have $\mathcal C\approx18$--$20$.

\section{Controlled variations}
\label{sec:app_controlled}

This appendix gives two additional tests used in the mechanism analysis of Sec.~\ref{sec:hierarchy}. Both use the same Salpeter wave functions as the main calculation.

\subsection{Localized $2S$ node-region remapping}
\label{sec:app_node_shift}

A monotonic piecewise-linear coordinate warp is applied locally to the $2S$ Salpeter wave function while leaving the $1S$ wave function unchanged. The warp is the identity at the edges of a $\pm0.25$~GeV window around the original node region and is constructed so that the target coordinate maps to the original zero, $h(q_{\rm target})=q_0$. The same coordinate warp is applied to all independent components, while $q_0$ is defined from the leading $S$-wave components: the mean zero of $f_5,f_6$ ($0.85485$~GeV) for the vector $2S$ state and of $g_1,g_2$ ($0.84828$~GeV) for the pseudoscalar. The resulting form factors are:
\begin{equation}
\begin{array}{c|ccc}
q_{\rm node}^{(S)}\,\text{(GeV)} & 0.70 & \text{original}\simeq0.85 & 1.00\\
\hline
|F_{\rm C3}|\,\text{(GeV}^{-1}\text{)} & 0.0175 & 0.0105 & 0.0390\\
|F_{\rm C4}|\,\text{(GeV}^{-1}\text{)} & 0.0163 & 0.0120 & 0.0432
\end{array}
\end{equation}
The remap explicitly reproduces the target zeros of the leading components. Relative to the unmodified wave functions, shifting the leading-$S$ zero to $0.70$~GeV enhances the C3 and C4 form factors by factors of about $1.7$ and $1.4$, whereas shifting it to $1.00$~GeV enhances them by factors of about $3.7$ and $3.6$. These values are used in the node-region sensitivity discussion of Sec.~\ref{sec:node_region}.

\subsection{Symmetrized recoil routing}
\label{sec:app_alpha_swap}

For comparison, setting $\alpha_1=\alpha_2=0.5$ in the current while keeping the masses and Salpeter wave functions unchanged gives:
\begin{equation}
\begin{array}{c|cc}
 & |J_b|/|J_{\bar c}|\text{ (C4)} & |J_b|/|J_{\bar c}|\text{ (C3)}\\
\hline
\alpha_1\neq\alpha_2\text{ (physical routing)} & 0.020 & 0.20\\
\alpha_1=\alpha_2=0.5\text{ (symmetrized)} & 0.045 & 0.065
\end{array}
\end{equation}
The resulting change in the C3/C4 line ratios is the input used in the routing discussion of Sec.~\ref{sec:quarkline_main}.

\end{document}